\documentclass{article}

\usepackage{amssymb}
\usepackage{amsmath}
\usepackage[numbers,sort&compress]{natbib}
\usepackage{bibunits}
\usepackage{graphicx}
\usepackage{siunitx}
\usepackage{xcolor}
\usepackage{hyperref}
\usepackage{comment}
\usepackage{caption}
\usepackage{subcaption}
\usepackage[thinc]{esdiff}
\usepackage[ruled]{algorithm2e}
\usepackage{booktabs}
\usepackage{authblk}
\usepackage{placeins}

\newcommand{\email}[1]{\href{mailto:#1}{\nolinkurl{#1}}}
\newcommand*{\doi}[1]{\href{https://doi.org/\detokenize{#1}}{doi: \nolinkurl{#1}}}

\begin{document}

\title{Continuation methods as a tool for parameter inference in electrophysiology modeling}

\author[1]{Matt J. Owen}
\affil[1]{Centre for Mathematical Medicine and Biology, School of Mathematical Sciences, University of Nottingham, University Park, Nottingham, NG7 2RD, United Kingdom}
\author[1]{Gary R. Mirams\footnote{Corresponding author: \email{gary.mirams@nottingham.ac.uk}}}
\date{}

\maketitle

\begin{abstract}
Parameterizing mathematical models of biological systems often requires fitting to stable periodic data. 
In cardiac electrophysiology this typically requires converging to a stable action potential through long simulations. 
We explore this problem through the theory of dynamical systems, bifurcation analysis and continuation methods; under which a converged action potential is a stable limit cycle. 
Various attempts have been made to improve the efficiency of identifying these limit cycles, with limited success.
We demonstrate that continuation methods can more efficiently infer the converged action potential as proposed model parameter sets change during optimization or inference routines. 
In an example electrophysiology model this reduces parameter inference computation time by 70\%. 
We also discuss theoretical considerations and limitations of continuation method use in place of time-consuming model simulations.
The application of continuation methods allows more robust optimization by making extra runs from multiple starting locations computationally tractable, and facilitates the application of inference methods such as Markov Chain Monte Carlo to gain more information on the plausible parameter space.
\end{abstract}

\section{Introduction}

It is a common task when modeling nonlinear biological systems to calibrate a model to periodic data. 
Such data arise naturally for a diverse range of phenomena: predator-prey population cycles \citep{jost2001pattern}, circadian rhythms \citep{asgari2019mathematical}, intracellular calcium oscillations \citep{han2017mathematical}, respiration \citep{mackey1977oscillation} and oscillations in neural activity \citep{donoghue2022methodological} to name but a few.
In this article we propose the use of continuation methods as part of parameter optimization and inference for such settings.
Below we take the example of calibration of a cardiac action potential model to apply continuation methods, and compare them with the standard approach of numerically solving differential equations for an extended period of time.

\subsection{Optimization of cardiac action potential models}
Mathematical models of cardiomyocyte electrophysiology typically comprise a system of Ordinary Differential Equations (ODEs) describing the proportion of ion channels and their conformational states in the cell membrane, and ion concentrations inside sub-cellular compartments, at a given time. 
To parameterize these models, it is typical to compare model simulated transmembrane-voltage waveforms (action potentials, APs), for a given set of conductance parameters, against experimentally derived APs \citep{tomekDevelopmentCalibrationValidation2019, duttaOptimizationSilicoCardiac2017, duInSilicoModelingGlycosylation2014, houstonReducingComplexityUnidentifiability2020,whittakerCalibrationIonicCellular2020}.

\begin{figure}[htb]
    \centering
    \includegraphics[width=\textwidth]{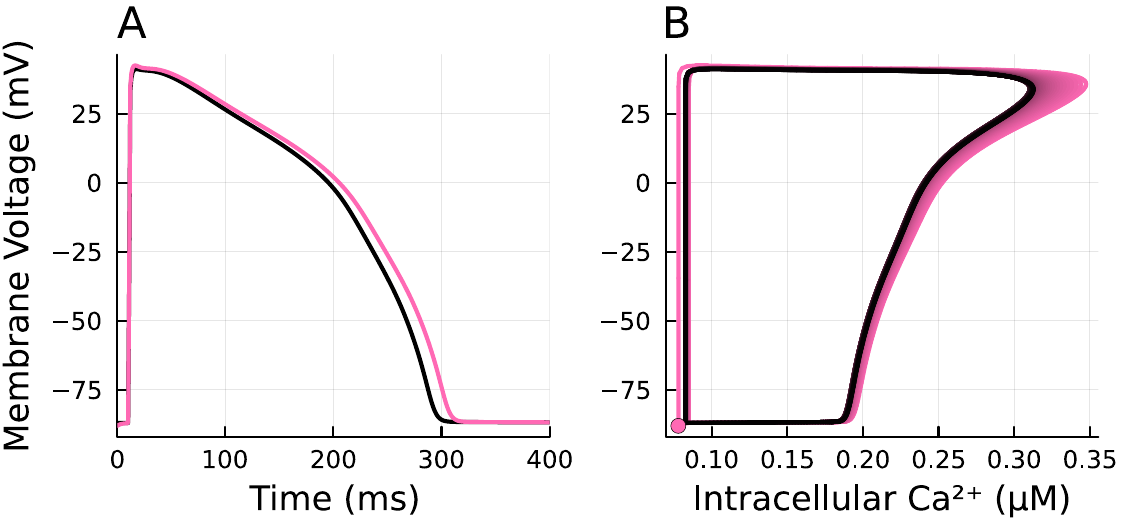}
    \caption{Convergence to a limit cycle or `stable action potential'. 
    Simulations used the O'Hara-Rudy CiPA model of \citet{duttaOptimizationSilicoCardiac2017}. 
    A: Two action potentials at different stages of convergence. 
    An unconverged AP in pink, and a more converged AP (after \qty{30}{\second}) in black. 
    B: Phase diagram of intracellular calcium concentration against membrane voltage, showing the convergence from a perturbed state (pink) to a stable action potential (black).
    }
    \label{fig:actionPotential}
\end{figure}

In single cell models, or isolated myocyte experiments, APs are typically initiated with an externally-applied stimulus current (`\emph{paced}'). 
This stimulus current perturbs the voltage across the cell membrane from its resting polarized state ($V\approx\qty{-80}{\mV}$), which then activates sodium-specific ion channels leading to a large fast sodium current to complete the depolarization of the cell (Fig~\ref{fig:actionPotential}A).
The cell then gradually repolarizes through a range of ion currents (predominantly potassium-specific) to return to its resting polarized state. 

This stimulus current is periodically applied to trigger APs. 
However, these ion currents cause changes in the concentrations of each ion within the cell, which impacts the future dynamics of ion channels and therefore future APs.
The pacing of the cardiomyocyte continues as the AP stabilizes, with all variables undergoing periodic orbits (Fig~\ref{fig:actionPotential}B).
The experimental AP for model parameterization is generally recorded after such stabilization has occurred.
Typically, the recorded AP consists of time-series data on membrane voltage, intracellular calcium concentration, or both \citep{whittakerCalibrationIonicCellular2020}.
The stabilized AP is dependent on many things, including the pacing frequency, so this stabilization process may need to be completed multiple times depending on the data being recorded.

During optimization, model simulations reproduce this process, running the ODE model until all states converge to a stable periodic orbit --- a limit cycle (typically between \qty{100}~ and \qty{1000}~ stimulus pulses at a frequency of \qty{1}{\Hz} \citep{duInSilicoModelingGlycosylation2014, clerxImmediateDelayedResponse2021, tomekDevelopmentCalibrationValidation2019, duttaOptimizationSilicoCardiac2017, smirnovGeneticAlgorithmbasedPersonalized2020, groenendaalCellSpecificCardiacElectrophysiology2015, houstonReducingComplexityUnidentifiability2020}).

The convergence to a limit cycle is slow which then leads to a large computational expense in evaluating a cost function at a given set of parameters, with each stabilized limit cycle requiring up to a thousand times the computational cost of a single orbit\footnote{This cost depends on the complexity of the model and the number of slowly changing states --- typically slow behavior is exhibited in models with concentrations in various compartments which slowly alter across many action potentials. Concentrations change as any small imbalance of net charge flow causes the system to evolve to a new stable state where net charge flow during a whole limit cycle is zero}.
Additionally, these intermediate orbits must be solved accurately as the change in the concentrations across an orbit is dependent on the dynamics during that specific orbit.

The effects of the slow convergence are then further compounded when model outputs are evaluated against action potential recordings from multiple different stimulus pacing frequencies \citep{syedAtrialCellAction2005}.

The highly fast-slow dynamics (fast upstroke of the AP and ion channel dynamics with slowly changing concentrations) and the high computational expense exhibited by these models has resulted in a range of attempts to better identify stabilized APs to perform model parameterization.
\citet{smirnovGeneticAlgorithmbasedPersonalized2020} included the initial ionic concentrations as parameters in a genetic algorithm, undergoing selection, crossover, and mutation. 
However, the next generation inherit from the final ionic concentrations.
This ensures the genetic algorithm converges to a stabilized AP but allows for simulating fewer stimulus pulses as the model converges over the course of the parameterization rather than just a single simulation.
As such, both the accuracy of the cost function and the dimensionality of the parameter space were sacrificed for reducing the computation time per AP.
This approach does not conserve $\Gamma_0$, a parameter measuring the difference in unmodeled ions inside and outside the cell membrane \citep{barralParameterRepresentingMissing2022}. 
Parameterization can be made more challenging when $\Gamma_0$ varies as it increases the dimensionality of the problem.

Other attempts to improve the optimization efficiency for these models include: 
\begin{itemize}
    \item fast approximations of exponentials and power calculations \citep{botRapidGeneticAlgorithm2012}; 
    \item reducing the number of limit cycle orbits for convergence during optimization, as an approximation of the fully converged limit cycle \citep{tomekDevelopmentCalibrationValidation2019}, although this can cause accuracy issues during the later stages of optimization if the limit cycle is not fully converged;
    \item utilizing data recorded starting from a resting steady state without a stimulus (stable equilibrium, fast to converge), as opposed to a paced steady state (stable limit cycle, slow to converge), or clamped ionic concentrations, which cause more rapid convergence \citep{faberActionPotentialContractility2000};
    \item and forgoing typical parameterization methods in favor of inferring model conductances through biomarker data by linear regression \citep{sarkarRegressionAnalysisConstraining2010}.
\end{itemize}

\subsection{Continuation methods}
Continuation methods provide a tool to explore how equilibria and periodic orbits such as limit cycles change as functions of model parameters.
When studying equilibria, for a generic ODE system given by Eq.~\eqref{equ:dynamicalSystem}, continuation methods aim to identify solutions to Eq.~\eqref{equ:equilibria}.
\begin{align}
    \diff{y}{t}&=F(y,\theta),
    \label{equ:dynamicalSystem}\\
    F(y,\theta)&=0.
    \label{equ:equilibria}
\end{align}
In particular, for a known solution $(y_0,\theta_0)$, they seek to identify the parametric curve $\gamma(s)=(y(s),\theta(s))$ where all its points are solutions to Eq.~\eqref{equ:equilibria} and the curve passes through the point $(y_0,\theta_0)$.
A common method for computing this curve involves predictor-corrector methods. 
These first identify the tangent of $\gamma(s)$ at a current point $(y^*,\theta^*)$, $(dy^*, d\theta^*)$, either numerically or analytically.
They then take a step in the tangent direction called the `predictor', $(y^*+dy^*, \theta^*+d\theta^*)$.
This is followed by a `corrector' step, where this approximate point is moved to a nearby solution, using a method such as the Newton root finding algorithm.
Additionally, some restriction needs to be placed on the corrector step to ensure the method does not return to $(y^*,\theta^*)$.
This process is repeated to travel along the curve of equilibria as a function of parameters in the $\pm s$ directions.

When working with periodic orbits, as opposed to equilibria, Eq.~\ref{equ:equilibria} is no longer valid.
We instead have that the solution should be periodic, $y(0)=y(T)$, for some period $T$.
If the flow solution of the ODE system is $\Phi(y_0,t,\theta)$ (i.e.\ $\Phi(y_0,t,\theta)$ describes the solution of Eq.~\eqref{equ:dynamicalSystem} that passes though $y_0$ as a function of time $t$), then this periodic restriction can be expressed as given by Eq.~\eqref{equ:flow}.
Since any point $y$ on the limit cycle will satisfy Eq.~\eqref{equ:flow}, it is necessary to include some phase condition for uniqueness (Eq.~\eqref{equ:phase}).
\begin{align}
    \Phi(y,T,\theta)-y&=0, \label{equ:flow} \\
    \varphi(y,T,\theta)&=0. \label{equ:phase}
\end{align}
Finally, the system of Eqs.~(\ref{equ:flow},~\ref{equ:phase}) can be solved using the predictor-corrector methods above. 
The flow solution $\Phi$ (the full periodic orbit solution for each variable) can be computed numerically for a single orbit using an ODE solver of choice.
While we described `shooting method' continuation above \citep{umbriaNumericalContinuationMethods2016}, other methods to perform continuation of periodic orbits are possible \cite{govaertsNumericalBifurcationAnalysis2000, dankowiczRecipesContinuation2013}. 

A wide range of implementations of continuation methods are available including \texttt{MatCont} \citep{dhoogeNewFeaturesSoftware2008}, \texttt{PyDSTool} \citep{clewleyPyDSToolSoftwareEnvironment2007}, \texttt{BifurcationKit.jl} \citep{veltzBifurcationKitJl2020}, \texttt{CoCo} \citep{dankowiczRecipesContinuation2013} and \texttt{AUTO} \citep{doedelAUTOProgramAutomatic1981}. 
We refer the interested reader to the following works for a comparison of these tools \citep{dhoogeNewFeaturesSoftware2008, blythTutorialNumericalContinuation2020, meijerNumericalBifurcationAnalysis2011, govaertsInteractiveContinuationTools2007}. 
We will utilize \texttt{BifurcationKit.jl} for this work, as it features a range of methods for continuation of periodic orbits. 
Additionally, the Julia language, through \texttt{DifferentialEquations.jl}, includes a suite of efficient ODE solvers to compare continuation methods against \citep{rackauckas2017differentialequations}.

\subsection{Parameter inference}
Complex mathematical models rarely have all parameters pre-determined experimentally. 
As such, various inference methods are used to determine parameters from experimental data and model simulations.
Experimental data are typically noisy, which introduces uncertainty into inferred parameters and in cardiac modeling this has been quantified with a range of inference methods \citep{huGeneralizedPolynomialChaosbased2018, coveneyFittingTwoHuman2018, coveneySensitivityUncertaintyAnalysis2020, pathmanathanComprehensiveUncertaintyQuantification2019, pathmanathanDataDrivenUncertaintyQuantification2020, claytonAuditUncertaintyMultiscale2020}.

In a Bayesian setting, we begin by defining a prior, $\pi(\theta)$, on the parameters and a likelihood of observing the data given the parameters, $\mathcal{L}(\theta \mid y)$ (defined by some statistical noise model).
While we typically cannot calculate the posterior directly, we do have many methods to sample from the posterior distribution.
In the case of MCMC, we construct a Markov chain with a steady state distribution of the posterior distribution (for example, with the Metropolis Hastings algorithm \citep{hitchcockHistoryMetropolisHastingsAlgorithm2003}).

Optimization algorithms, and Bayesian inference methods in particular (such as Markov Chain Monte Carlo, MCMC \citep{haarioAdaptiveMetropolisAlgorithm2001}, or Approximate Bayesian Computation, ABC \citep{toniApproximateBayesianComputation2008}), require a large number of model evaluations.
Additionally, it is often beneficial to repeat optimizations from multiple starting locations.
For this reason, improvements in the speed at which the model outputs can be evaluated directly leads to improvements in the quality of optimization and inference, producing more reliably estimated parameters and more accurate quantification of their uncertainty.

\citet{NEURIPS2021_2c6ae45a} recently used continuation and parameter inference methods together to reproduce user-specified bifurcation diagrams where time series data are unavailable or impractical.

In this work, we investigate the use of continuation methods as tools for tracking the trajectory of model APs as new parameter sets are investigated.
We demonstrate how continuation can reduce computational cost, for computing converged APs, but also more generally across an entire model inference exercise though using continuation within an MCMC scheme.
Finally, we explore the theoretical consequences that may arise from using continuation methods in more complex models.

While we focus our applications towards cardiac AP models, the methods outlined here apply more generally to fast-slow dynamical systems, where the data against which they are optimized is derived from some kind of stable steady state (stable equilibria or limit cycles). 

\section{Model}
We first define a toy cardiac AP model, which we derive from the Nobel 1962 model \citep{nobleModificationHodgkinHuxleyEquations1962}.
In its original formulation, this model, given by Eq.~\eqref{equ:noble}, contains four states, three of which are ion channels and one of which is the membrane voltage. 
For simplicity we restrict ourselves to consider variation in only three parameters, scaling factors for the sodium, potassium, and leak conductances, $\theta_1$, $\theta_2$, and $\theta_3$, respectively.

The Noble 1962 model \citep{nobleModificationHodgkinHuxleyEquations1962} is for Purkinje fibres and produces APs spontaneously without a stimulus current.
This means we additionally need to consider the changing period of the APs, which is typically defined by the stimulus frequency in models for the atria or ventricles. 
However, this simplifies its implementation within \texttt{BifurcationKit.jl} as currently it does not support non-autonomous ODEs (e.g.\ models with a stimulus current applied at particular times). Although these systems can be included if the model can be transformed into an autonomous system (for example, by augmenting the system with additional state variables to form a harmonic oscillator with the desired period, then defining a stimulus in terms of one of these new variables).

\subsection{Including concentrations}
The Noble 1962 model does not include concentrations, which are typically the slowest variables in fast-slow AP models.
As such, we have derived a new model from it which includes sodium and potassium concentrations.
Rather than including the additional currents that would normally be required to include these concentrations into the model, such as the NaK-ATPase pump, we include general forcing terms that push the concentrations towards target concentrations $[\tilde{\text{Na}}]$ and $[\tilde{\text{K}}]$ which produce fixed target reversal potentials $\tilde{E}_\text{Na}$ and $\tilde{E}_\text{K}$, respectively.
We also include a scaling parameter $\tau$ on the rate of decay to these target reversal potentials, which we will use to control the limit cycle convergence rate to create an exemplar model with typical cardiac AP model convergence properties.

A simplified form of this model is given by Eq.~\eqref{equ:simpleNobleConc} while the full set of equations is given in Eq.~\ref{equ:nobleConc}

\begin{equation}
\begin{aligned}
\diff{V}{t} &= -\frac{i_\text{Na}+i_\text{K}+i_\text{leak}}{12},&
i_\text{Na} &= (g_\text{Na}+140)(V-E_\text{Na}([\text{Na}]_i)),\\
\diff{m}{t} &= \alpha_m(V)(1-m)-\beta_m(V)m,&
i_\text{K} &= (g_\text{K1}+g_\text{K2})(V-E_\text{K}([\text{K}]_i)),\\
\diff{h}{t} &= \alpha_h(V)(1-h)-\beta_h(V)h,&
i_\text{leak} &= 75\times\theta_3(V+60),\\
\diff{n}{t} &= \alpha_n(V)(1-n)-\beta_n(V)n,&
g_\text{Na} &= 400000\times\theta_1m^3h,\\
g_\text{K1} &= 1200\times\theta_2f(V)+g(V),&
g_\text{K2} &= 1200\times\theta_2n^4,\\
[\tilde{\text{Na}}]_i &= [\text{Na}]_o\exp\left(-\tilde{E}_\text{Na}\frac{F}{RT}\right),&
[\tilde{\text{K}}]_i &= [\text{K}]_o\exp\left(-\tilde{E}_\text{K}\frac{F}{RT}\right),\\
\diff{[\text{Na}]_i}{t} &= \tau\left(-\frac{i_\text{Na}}{1000F} - \frac{[\text{Na}]_i-[\tilde{\text{Na}}]_i}{20}\right),\span \span \\
\diff{[\text{K}]_i}{t} &= \tau\left(-\frac{i_\text{K}+i_\text{leak}}{1000F} - \frac{[\text{K}]_i-[\tilde{\text{K}}]_i}{20}\right). \span \span \label{equ:simpleNobleConc}
\end{aligned}
\end{equation}

\subsection{Convergence test}
\label{sec:convergenceTest}
To verify that a solution is converged, we evaluate the change in state variables every orbit. 
If the state variables return to similar values after one orbit, it is considered converged.

Specifically, every time the solution to the ODE passes $V=\qty{-20}{\mV}$ with $\diff{V}{t}>0$ (upcrossing), we determine the change in each state variable compared with the previous upcrossing of $V=\qty{-20}{\mV}$. 
We calculate the $L_1$ norm of the change in these states, and if the norm is below $10^{-6}$, we consider the limit cycle to have converged.
This method ($L_1$ norm of below $10^{-6}$ across one AP) is used to determine convergence in the cardiac simulation software Chaste \citep{cooperChasteCancerHeart2020, cooperCellularCardiacElectrophysiology2015}.

In addition to testing convergence as above, continuation methods in \texttt{BifurcationKit.jl} also include their own convergence condition\footnote{If the $L_2$ norm of the residual vector (error in states and period across one orbit) is less than $10^{-10}$, then \texttt{BifurcationKit.jl} considers the limit cycle to be converged.}.
In this work, continuation methods are required to meet both conditions. 
The use of a secondary convergence criteria for continuation methods favors the ODE methods in the following benchmarks, which only need to meet the one condition specified above.

\subsection{Model convergence rate}
To finalize the model, we define $\tau$ to ensure the model converges on a typical timescale. 
Most cardiac AP models converge within a few hundred seconds \citep{clerxImmediateDelayedResponse2021, smirnovGeneticAlgorithmbasedPersonalized2020}, but some models take over \qty{10000}{\second} \citep{clerxImmediateDelayedResponse2021}, although different works use slightly different definitions of convergence. 

When running optimizations, models are typically run for between \qty{100}{\second} and \qty{1000}{\second} for data of around 0.5--2\,Hz \citep{duInSilicoModelingGlycosylation2014, houstonReducingComplexityUnidentifiability2020, smirnovGeneticAlgorithmbasedPersonalized2020, duttaOptimizationSilicoCardiac2017, tomekDevelopmentCalibrationValidation2019}, although in some rare cases timescales over \qty{1000}{\second} may be used \citep{deckerPropertiesIonicMechanisms2009, groenendaalCellSpecificCardiacElectrophysiology2015}.

Since we desire a simple model that replicates these timescales of convergence, we set $\tau=2.5$. 
In this case, starting from the default model initial conditions in Table~\ref{tab:ic} (Non-converged) we converge in around \qty{100}{\second}.
Eq.~\ref{equ:nobleConc} gives the final Noble concentration model.

\section{Convergence approaches}
We evaluate three different approaches for identifying a converged limit cycle:
\begin{enumerate}
    \item the \emph{Standard Approach}, where the model is solved numerically with an ODE solver from a standard set of initial conditions given in Table~\ref{tab:ic} (Non-converged), using the proposed parameters. 
    \item the \emph{Tracking Approach}, where the model is solved numerically with an ODE solver from some previously converged limit cycle, using the proposed parameters.
    \item the \emph{Continuation approach}. Continuation is performed from some previous limit cycle (and its corresponding parameters) towards the proposed parameters (iteratively, possibly across multiple steps) to identify their corresponding limit cycle. 
    Here, we utilize shooting method continuation.
\end{enumerate}

For both ODE approaches, Standard and Tracking, we solve the model until convergence and then terminate (up to a maximum simulated time of \qty{10000}{\second}).

\section{Simulation times}
\label{sec:simulationTimes}
In this section, we demonstrate that continuation methods are capable of tracking limit cycles through parameter perturbations with greater efficiency than the presently-used ODE solvers. 
We then explore how this efficiency is influenced by the size of the perturbation.

\subsection{Small perturbation}
All three approaches, Standard, Tracking, and Continuation, are used to identify a converged limit cycle for $\theta_1=1.1, \theta_2=1,\theta_3=1$.
Where applicable, the approaches use a preconverged limit cycle from $\theta_1,\theta_2,\theta_3=1$.

The performance of each approach is evaluated using \texttt{BenchmarkTools.jl} \citep{chenRobustBenchmarkingNoisy2016} and the convergence of each approach is verified with the method described previously.

\begin{figure}[htb]
    \centering
    \includegraphics[width=\textwidth]{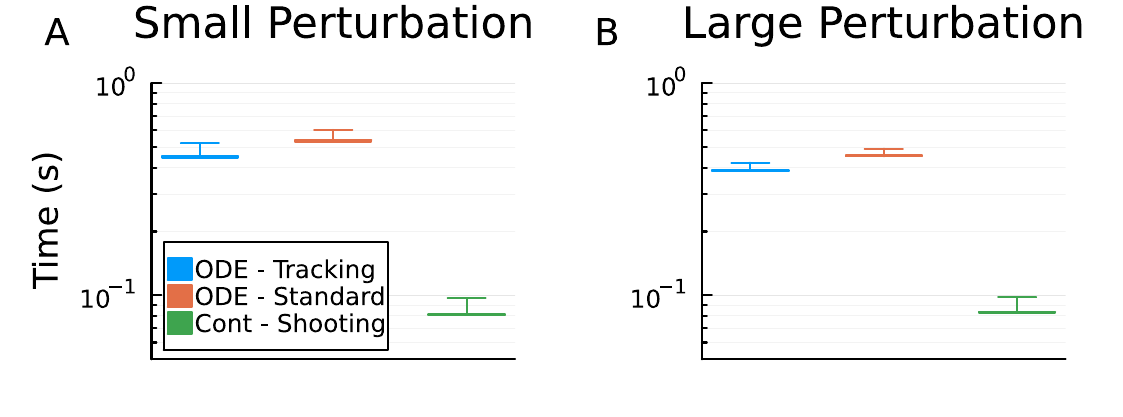}
    \caption{Benchmarking computation times for differently sized parameter perturbations. 
    Box plots show the quartiles across repeated simulations with whiskers are the maximum and minimum. The 0th, 25th, 50th, and 75th percentiles overlap.
    Computation times reported in seconds for: `ODE - Tracking' ODE simulation from previous limit cycle (Table~\ref{tab:ic} - Converged), `ODE - Standard' ODE simulation from standard non-converged initial conditions, `Cont - Shooting' Shooting method continuation to follow a previously converged limit cycle (Table~\ref{tab:ic} --- Converged). A: Small perturbation $(\theta_1=1, \theta_2=1, \theta_3=1)\rightarrow(\theta_1=1.1,\theta_2=1,\theta_3=1)$. B: Large perturbation $(\theta_1=1, \theta_2=1, \theta_3=1)\rightarrow(\theta_1=1.5,\theta_2=1.2,\theta_3=0.8)$
    }
    \label{fig:simTimings}
\end{figure}

The times to compute the converged limit cycle are given in Fig~\ref{fig:simTimings}A. 
Continuation significantly outperforms the Standard approach by a factor of $6.5$ and the Tracking approach by a factor of $5.5$. 
There is a slight reduction ($16\%$) in compute time by using the Tracking approach over the Standard approach.

Since the parameter perturbation is small, the Continuation approach is able to jump from the starting parameters to the final parameters and still converge to a limit cycle without requiring any intermediate iterations.

\subsection{Large perturbation}
We now repeat the same exercise for a larger parameter perturbation. 
This larger perturbation attempts to identify a converged limit cycle for $\theta_1=1.5, \theta_2=1.2, \theta_3=0.8$. 
Where applicable, the approaches again use a preconverged limit cycle from $\theta_1,\theta_2,\theta_3=1$.
Fig~\ref{fig:simTimings}B presents the computation times for each of the convergence approaches.

The larger parameter perturbation requires multiple iterations of continuation (performed automatically) in order to be stable, although this only gives a minor increase ($2\%$) in compute time compared with the small perturbation. 

While the Continuation approach exhibited an increase in computational cost, the Tracking and Standard approaches actually found this limit cycle even easier to converge to than the small parameter perturbation. 
This resulted in a slight decrease in computation time compared with the small perturbation ($13\%$ for Tracking, $15\%$ for Standard).

\begin{figure}[htb]
    \centering
    \includegraphics[width=\textwidth]{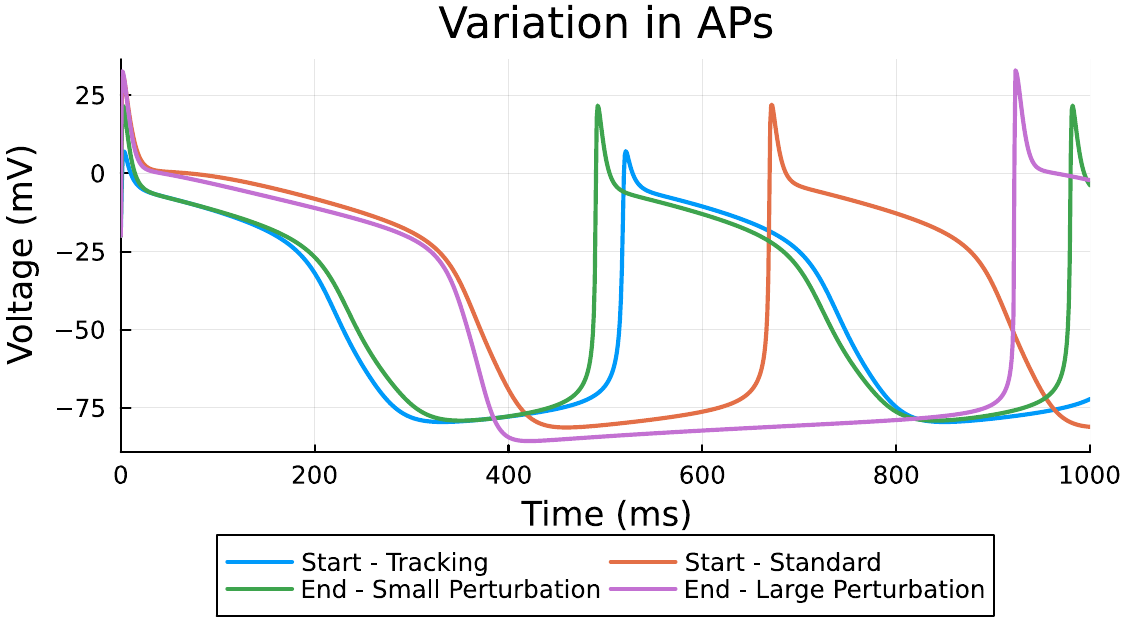}
    \caption{APs for the Standard and Tracking approaches, aligned on the first upcrossing of $V=\qty{-20}{\mV}$. `Start - Tracking' --- the converged AP for $\theta_1=1,\theta_2=1,\theta_3=1$ which is used to initiate the Tracking approach (Table~\ref{tab:ic}). `Start - Standard' --- the non-converged initial AP for the Standard approach (Table~\ref{tab:ic}). `End - Small Perturbation' --- the converged AP for the small perturbation parameters, $\theta_1=1.1,\theta_2=1,\theta_3=1$. `End - Large Perturbation' --- the converged AP for the large perturbation parameters, $\theta_1=1.5,\theta_2=1.2,\theta_3=0.8$.
    }
    \label{fig:simTimingAPs}
\end{figure}

Fig.~\ref{fig:simTimingAPs} shows the APs at the start of the Standard and Tracking approaches, and the converged APs for the small and large perturbed parameters. 
It is surprising that even the Tracking approach, which begins very close to the converged AP for the small perturbation, still converges to the AP for the large perturbation more quickly.

We begin to identify the cause of this discrepancy by noting that this decrease in computation time does not come from reaching convergence sooner (in simulated time), as shown by Table~\ref{tab:simTimingConvTime}. 
It instead appears to come from the increase in the period of the limit cycle at the large perturbation parameters.
This larger period allows the ODE solver to take larger time steps, which means that the computational expense of each orbit is similar, but fewer orbits are required to converge to the large perturbation parameters, and so convergence is computationally cheaper.

\begin{table}[htb]
\centering
\begin{tabular}{c c c}
\toprule
Approach & Small Perturbation & Large Perturbation \\
\midrule
Standard & 105 & 111 \\
Tracking & 88 & 95 \\
\bottomrule
\end{tabular}
\caption{Simulated time to reach convergence (in seconds) for the Standard and Tracking approaches.}
\label{tab:simTimingConvTime}
\end{table}

Continuation remains significantly quicker than the ODE approaches, although their reductions in cost for the larger perturbation make the differences smaller (for the Tracking approach, Continuation is $4.7$ times faster compared with $5.5$ for the small perturbation; for the Standard approach, Continuation is $5.5$ times faster compared with $6.5$ for the small perturbation).

The computational cost of continuation methods is dependent of the distance in parameter space and the sensitivity of the limit cycle (how much it changes for a given change in parameters).
In comparison, the Standard approach is independent of the change in both the parameters and the limit cycle (since it does not start from a converged state so has no corresponding starting parameters or starting limit cycle) and the Tracking approach is dependent only on the change in the limit cycle without any explicit dependence on the distance in parameter space.

\section{Parameter inference --- MCMC}
When identifying parameters in cardiac AP models, it can be beneficial to estimate the posterior distribution, not just due to the common occurrence of parameter unidentifiability in such models, but also to put accurate confidence bounds on identifiable parameters and identify multi-model posteriors.
Unfortunately, estimating the posterior distribution, as opposed to identifying a single set of parameters, further increases the computational expense in parameter inference.

MCMC samplers are a common tool for sampling posterior distributions for these models \citep{johnstoneHierarchicalBayesianModelling2016, johnstoneUncertaintyVariabilityModels2016, ramosQuantifyingDistributionsParameters2021}.
Additionally, MCMC schemes commonly propose small steps in parameter space, especially when in settings with good parameter identifiability, which are ideal for continuation methods where the computational cost increases with larger steps in parameter space.

\subsection{Methods}
\label{sec:mcmcMethods}
We have implemented an Adaptive Metropolis Hastings MCMC sampler \citet{haarioAdaptiveMetropolisAlgorithm2001} that has been previously applied to cardiac AP models \citet{johnstoneUncertaintyVariabilityModels2016}.

In addition to storing the last accepted state (conductance parameters and noise) of the MCMC and its likelihood, we also store a single point on the associated converged limit cycle.
When a new state is proposed, we calculate the converged limit cycle and its likelihood. 
If we accept this state, using a Metropolis Hastings acceptance probability, then in addition to updating the last accepted MCMC state, we also update the corresponding limit cycle and likelihood.

The Tracking and Continuation approaches will make use of the previously converged limit cycle from MCMC, while the Standard approach will always use the same initial conditions, given in Table~\ref{tab:ic} (Non-converged).

The Continuation approach always attempts to immediately jump the full parameter step, halving the step size if that fails to converge.

The data are generated by running the Standard approach to reach a stable limit cycle for $\theta_1,\theta_2,\theta_3=1$. 
From this limit cycle, we record a single orbit at time steps of \qty{1}{\ms}, starting and ending at an upcrossing of $V=\qty{-20}{\mV}$.
We include iid noise (standard deviation of \qty{2}{\mV}) on top of this data, which is then used for the MCMC of all approaches.
Further details of how the data are generated and how the model simulations are compared against them are given in Appendix~\ref{sup:mcmc}.

The MCMC algorithm begins at an initial guess found through maximum likelihood estimation using Nelder-Mead \citep{nelderSimplexMethodFunction1965} from \texttt{Optim.jl} \citep{mogensenOptimMathematicalOptimization2018} using the Standard approach. 

\subsection{Results}
The times to compute \qty{40000}~ MCMC iterations using each approach are given in Table \ref{tab:mcmcTimes}.
The Continuation approach provides a significant reduction in compute time (70\% reduction compared with the Standard approach). 
The estimated posterior distribution (after a 25\% burn in phase), from the Continuation approach, is shown in Fig~\ref{fig:posteriorCont}.
As shown in Appendix~\ref{sup:mcmc}, using continuation does not change the accuracy of the posterior distributions or how many iterations the MCMC takes to converge.
Additionally, it is slightly quicker to use the Tracking approach over the Standard approach ($7\%$ reduction). 

\begin{table}[htb]
\centering
\begin{tabular}{c c}
\toprule
Convergence approach & Compute time (hours) \\
\midrule
Standard & 12.41 \\
Tracking & 11.59 \\
Continuation & 3.71 \\
\bottomrule
\end{tabular}
\caption{The times to compute \qty{40000}~ iterations of MCMC with each approach.
The computation was performed on a single core using the University of Nottingham's High Performance Compute service Ada.
}
\label{tab:mcmcTimes}
\end{table}

\begin{figure}[htbp]
     \centering
     \includegraphics[width=\textwidth]{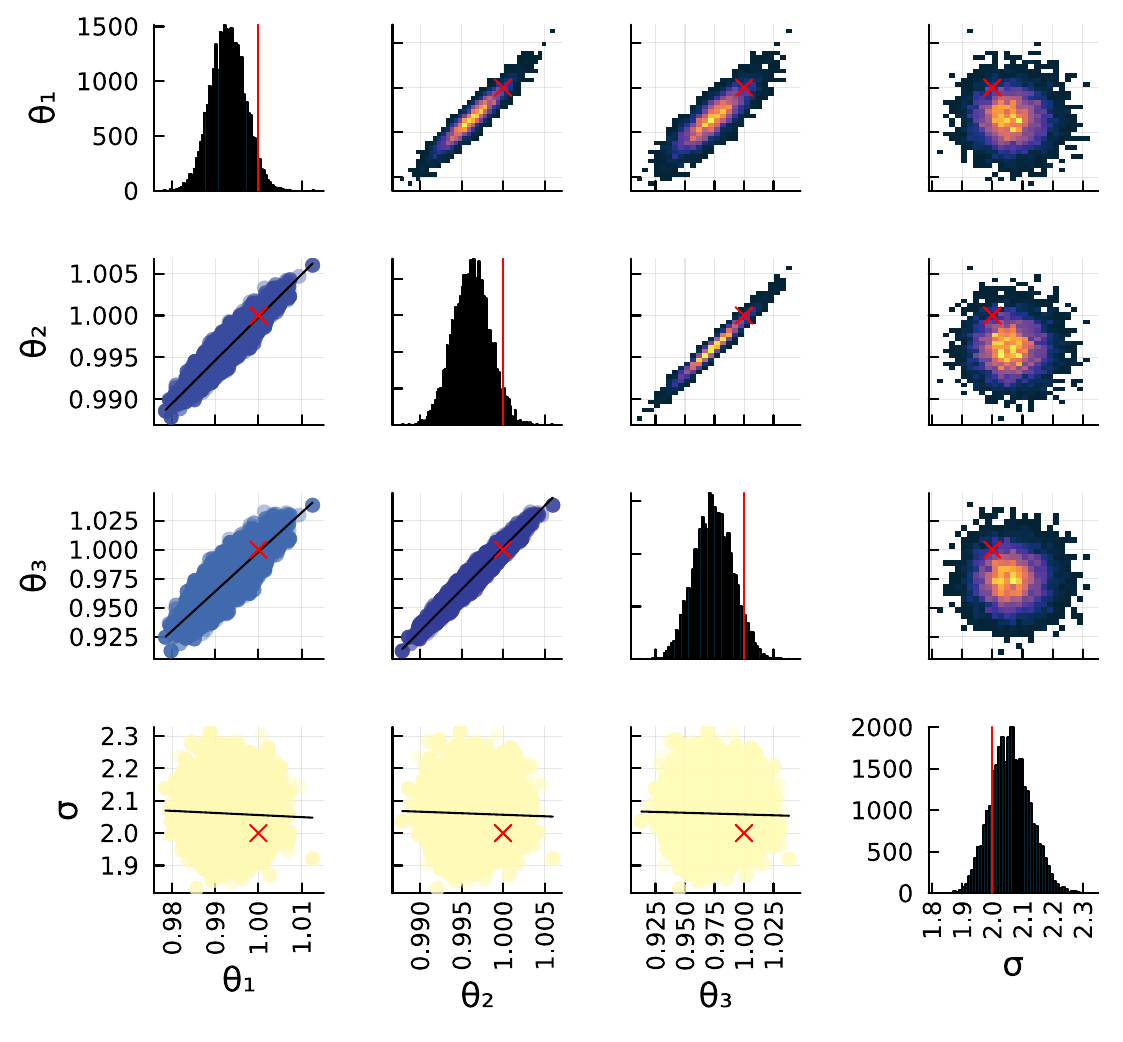}
    \caption{MCMC estimated posterior. The Noble concentration model, Eq.~\ref{equ:nobleConc}, is fitted to a single noisy converged AP from the data generating parameters $\theta_1=1, \theta_2=1,\theta_3=1$. The method for generating the data and calculating the likelihood are summarized in Section \ref{sec:mcmcMethods} with further details given in Appendix~\ref{sup:mcmc}. The lower triangle shows scatter plots (blue: strong correlation, yellow: weak correlation). The upper triangle shows heatmaps of the bivariate densities. The diagonal contains histograms of the marginal densities. The data generating parameters are given in red. Results shown are from the Continuation approach.
    }
    \label{fig:posteriorCont}
\end{figure}

\section{Implications of using continuation methods}
In this section, we will discuss some of the limitations with using continuation methods for these models.
An aspect of cardiac AP models that we have yet to address is the existence of multiple stable limit cycles for a particular parameter set \citep{surovyatkina2010multistability}.

\subsection{Duplicate limit cycles}
\begin{figure}[htb]
    \centering
    \includegraphics[width=\textwidth]{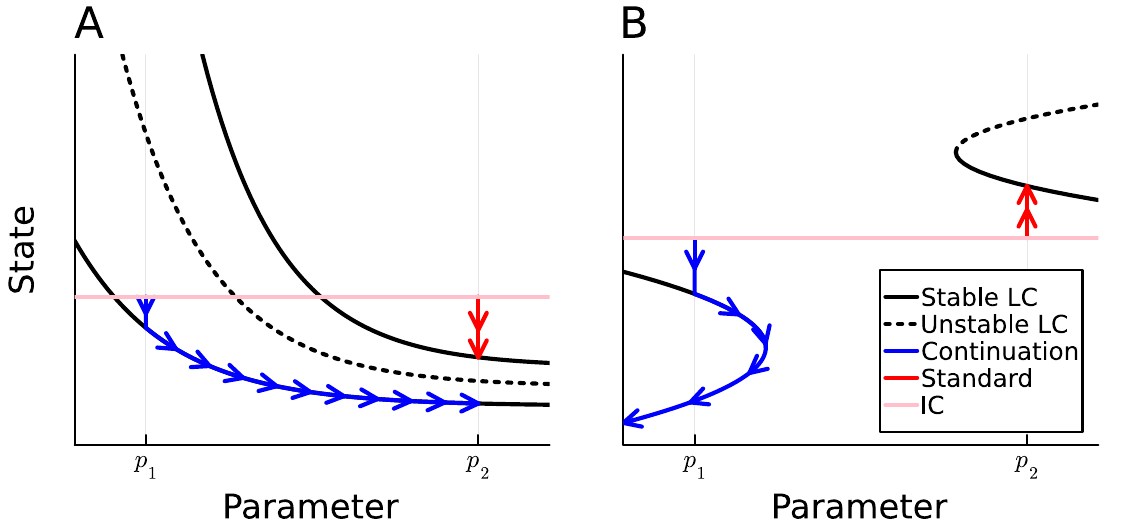}
    \caption{Examples of possible problems with using continuation methods when multiple limit cycles are possible. 
    A: Different methods can converge to different limit cycles. 
    If the Continuation approach initially converges to the limit cycle at $p_1$ (starting from the pink initial condition line), and then follows this limit cycle to $p_2$ (blue trajectory), it identifies a different limit cycle to the Standard approach which is run from the initial conditions at $p_2$ (red trajectory). 
    B: The limit cycle at $p_1$ may not exist at $p_2$. The Continuation approach (blue trajectory) finds a limit cycle at $p_1$ which it attempts to follow towards $p_2$, but the limit cycle doesn't exist at $p_2$. 
    As such, the Continuation approach instead follows the unstable branch, while this cannot happen to the Standard approach (red trajectory). 
    LC: Limit cycle, IC: Initial condition.}
    \label{fig:limitCycleProblems}
\end{figure}

Fig~\ref{fig:limitCycleProblems}A highlights how it would be possible to converge to a different limit cycle depending on the chosen method.
Here, the Continuation approach converges to a limit cycle at parameter vector $p_1$, and then follows this limit cycle to parameter vector $p_2$. 
However, the Standard approach converges from the same initial conditions but starting at parameter $p_2$, which leads to the two approaches converging to different limit cycles.

If the Continuation and Standard approaches do identify different limit cycles, and the initial conditions of the Standard approach are representative of the initial conditions used in the data generating process (initial conditions for the cardiomyocyte experimentally), it is more likely that the Standard approach is the one to identify the `correct' limit cycle.
Although, it is unlikely that the initial conditions used in the Standard approach are sufficiently accurate enough for this to be reliable.

\subsection{Extinction of limit cycles}
Another possibility is highlighted in Fig~\ref{fig:limitCycleProblems}B. 
Here, the limit cycle to which the Continuation approach converged at parameter $p_1$ does not exist at parameter $p_2$. 
However, another limit cycle does exist at parameter $p_2$, that we discover using the Standard approach. 
The continuation methods will track the $p_1$ limit cycle, even after it becomes unstable.

Pragmatically speaking, it is unlikely for a limit cycle to vanish completely in cardiac AP models that utilize a stimulus current, as the inclusion of a periodic forcing term almost always produces limit cycles.

\section{Conclusions}
In this work, we have proposed a new method for converging cardiac AP models utilizing continuation methods from dynamical systems theory. 
We have shown that these methods can reduce the time to converge a model to a stable limit cycle in some instances, and can do so reliably enough that they can reduce the time to run an MCMC algorithm by $70\%$.

While there are significant gains that can be made from these methods, there are also problems, both theoretical and practical that should be addressed.
From a practical standpoint, these methods can be challenging to implement and apply to more complex models than the one presented here. 

All ion-conserving cardiac electrophysiology models that we know of can converge to a limit cycle \citep{livshitz2009uniqueness}, although this is sometimes of period greater than one --- for example in the case of alternans or other higher-period behavior. 
Some models can respond chaotically at very fast pacing rates \citep{surovyatkina2010multistability}. 
One assumption here is that limit cycle convergence is desirable behavior, and a feature of the real biological systems. 
The extent to which this is true is something of an open question. 
At the whole heart level there are certainly indications of chaotic behavior \citep{babloyantz1988normal};
but this may be predominantly due to higher-level physiological processes are associated with feedback controls which constantly adapt heart rate and contractile force to maintain blood pressure and meet metabolic demands. 
At the isolated cardiac cell level, there is also beat-to-beat variability observed in the action potential \citep{zaniboni2000beat}, at least partly due to stochastic ion channel gating \citep{heijman2013determinants}. 
The fact beat-to-beat variability decreases substantially in tissue due to electrotonic coupling may mean that calibrating a cell model at a limit cycle may still result in a `good model' for a cell in tissue.
Still, even if calibrating deterministic limit cycle model behaviour to chaotic data is not ideal, one needs to choose some state for this class of models when undertaking calibration.
The model limit cycle therefore remains the most intuitive choice given that initial conditions for all variables are extremely difficult or impossible to measure simultaneously, and so the importance of minimizing computation when calculating the limit cycle remains.

Cardiac action potential models are particularly stiff, and this can prove problematic for the simple Newton optimisers used in the continuation methods, which results in them failing to converge to a stable limit cycle.
From a theoretical standpoint, identifying a stable limit cycle may not be sufficient if there are many stable limit cycles, and continuing from a `correct' previous limit cycle provides no guarantees that the later derived limit cycles will also be `correct'.

In some cases, the multiple limit cycles are due to differences in the value of $\Gamma_0$ (the difference between unmodeled intracellular and extracellular charges) across different initial conditions \citep{barralParameterRepresentingMissing2022}. 
These additional limit cycles (those with different values of $\Gamma_0$) can be safely ignored if this conserved quantity is known.
There may be other methods or conserved quantities that could simplify the limit cycle landscape of these models.

Ensuring continuation methods converge for the larger AP models in current use, such as \citet{duttaOptimizationSilicoCardiac2017} or \citet{tomekDevelopmentCalibrationValidation2019}, can be challenging.
As such, tracking the limit cycle throughout an MCMC optimization but still converging with an ODE solver may be a good compromise (Tracking vs.~Standard).
Using this approach still provides a small speed-up (7\%) with little increase in computational complexity.

It is also important to highlight the benefit of terminating the ODE approaches at convergence rather than running for a fixed amount of simulated time (see Section \ref{sec:convergenceTest}). 
If the period is fixed (typically the case for AP models, but not for the model used here), the convergence of the limit cycle can be easily obtained by comparing the state at the start of one AP with the state at the start of the next (for example, using the $L_1$ norm). 
However, even if the period is not fixed, utilizing ODE event handling (as done here) grants a cheap method for evaluating convergence of the model.
Terminating the ODE approaches once convergence is detected allows the ODE solver to only compute the orbits that are required for convergence, avoiding any excess computation.

The fact that continuation methods may not always identify the same limit cycle as the Standard approach means they present similarly to emulators, producing fast approximations of an underlying `true' solution\footnote{Although the Standard approach may still not identify the `correct' limit cycle in terms of the one giving rise to the biological data.}.
However, emulators typically require a large upfront training cost (not required here), and introduce additional uncertainty.
Even when it is preferable to use an emulator, it may be beneficial to use continuation methods to define the training data for the emulator, allowing more data to be included in the training set (for a given computational cost) over the Standard approach.

Although it is not ideal that different convergence methods may identify different limit cycles, this is not a drawback of using continuation methods.
While continuation methods highlight the existence of this problem, all approaches (including the Standard approach) suffer from this problem as there is currently no way to determine the `correct' limit cycle.
Continuation methods may actually provide a better strategy for handling the existence of multiple limit cycles than other approaches. 
They provide a mechanism to follow and perform continuation starting from multiple limit cycles (aided by their reduced computational cost for convergence), which could be incorporated into future inference methods.

While we have highlighted the benefits of continuation methods for optimizations with short step sizes, such as MCMC, these methods may also benefit other optimization algorithms.
Rather than only storing the last accepted limit cycle, we could store all previously identified limit cycles.
Then, to identify the limit cycle of a new set of parameters, we could chose the limit cycle of the nearest parameters that we have previously identified. 
This would allow the same methods to be used for any optimization algorithm.

The convergence of the Noble concentration model appears to be dependent on simulated time rather than number of orbits (see Section \ref{sec:simulationTimes}).
This is in contrast to cardiomyocytes which converge more quickly as pacing frequency increases \citep{pueyoMechanismsVentricularRate2010}, and therefore are dependent on the number of orbits rather than time.
We speculate that models can exhibit either behavior based on convergence being limited by sodium (and featuring only fast sodium currents active once per AP) or potassium (which passes through various channels which are active throughout depolarized and polarized potentials).
Depending on the specific choice of convergence criteria, specifically how closely each concentration is required to be converged, will alter whether the convergence of a model is limited by sodium or potassium.

Improving the convergence rate of AP model computations will allow more complex parameter inference to be utilized, such as the Bayesian methods shown here, population of models approaches \citep{muszkiewiczVariabilityCardiacElectrophysiology2016}, higher dimension inference (such as inferring some rate parameters instead of only conductances), or optimal experimental design \citep{shuttleworthEmpiricalQuantificationPredictive2023}.
We believe efficient parameter inference will become increasingly critical as these tools are used in emerging areas such as personalized medicine \citep{lopez-perezPersonalizedCardiacComputational2019}.

\section*{Funding}
This work was supported by the Wellcome Trust (grant no. 212203/Z/18/Z). 
GRM and MJO acknowledge support from the Wellcome Trust via a Wellcome Trust Senior Research Fellowship to GRM. 

This research was funded in whole, or in part, by the Wellcome Trust [212203/Z/18/Z]. 

\section*{Data availability}
The code and data generated in this work can be found on GitHub: \url{https://github.com/CardiacModelling/ContinuationForOptimisation}. 
An archived version of the code and generated data at time of publication is available on Zenodo (\url{https://doi.org/10.5281/zenodo.14623619}).

\section{Acknowledgments}
We are grateful for access to the University of Nottingham's Ada HPC service.\\


\renewcommand{\thetable}{\Alph{section}\arabic{table}}
\renewcommand{\thefigure}{\Alph{section}\arabic{figure}}
\renewcommand{\theequation}{\Alph{section}\arabic{equation}}

\appendix


\section{Appendix: Model}
The original Noble model is given by Eq.~\eqref{equ:noble}. 
The complete model, including concentrations, is given by Eq.~\eqref{equ:nobleConc}.

The default model initial conditions, used for defining the convergence rate $\tau$ and the Standard approach, are reported in Table~\ref{tab:ic} (Non-converged).
For the voltage $V$ and ion channel state variables $m$, $h$, $n$, they are defined as given in \citet{nobleModificationHodgkinHuxleyEquations1962}. 
The concentrations, $[\text{Na}]_i$ and $[\text{K}]_i$, are set to give approximately $E_\text{Na}=\qty{40}{\mV}$ and $E_\text{K} = \qty{-100}{\mV}$ (the values used in \citet{nobleModificationHodgkinHuxleyEquations1962}), while maintaining close to realistic values for the intracellular and extracellular concentrations. 
This table also gives initial conditions used for the $(\theta_1=1, \theta_2=1, \theta_3=1)$ limit cycle.

The model is solved using \texttt{DifferentialEquations.jl} \citep{rackauckas2017differentialequations} with the \texttt{Tsit5} ODE solver and an absolute tolerance of $10^{-10}$ and relative tolerance of $10^{-8}$.

\begin{table}[htb]
\centering
\begin{tabular}{c c c}
\toprule
Variable & Non-converged & Converged \\
\midrule
$V$ & \qty{-87}{\mV} & \qty{-79.090}{\mV} \\
$m$ & 0.01 & 0.049845 \\
$h$ & 0.8 & 0.80840 \\
$n$ & 0.01 & 0.55027 \\
$[\text{Na}]_i$ & \qty{30}{\milli M} & \qty{36.957}{\milli M} \\
$[\text{K}]_i$ & \qty{160}{\milli M} & \qty{153.78}{\milli M} \\
\bottomrule
\end{tabular}
\caption{The initial conditions used for the model. `Non-converged' is used as the initial condition for the Standard approach. `Converged' gives a point on the $(\theta_1=1, \theta_2=1, \theta_3=1)$ limit cycle which is used for the starting limit cycle in Section 4 for the Tracking and Continuation approaches.}
\label{tab:ic}
\end{table}

\begin{equation}
\begin{aligned}
\diff{V}{t} &= -\frac{i_\text{Na}+i_\text{K}+i_\text{leak}}{12}, &
\diff{m}{t} &= \alpha_m(1-m)-\beta_mm,\\
\diff{h}{t} &= \alpha_h(1-h)-\beta_hh, &
\diff{n}{t} &= \alpha_n(1-n)-\beta_nn, \\
\alpha_m &= \frac{-100(V+48)}{\exp\left(-\frac{V+48}{15}\right)-1}, &
\beta_m &= \frac{120(V+8)}{\exp\left(\frac{V+8}{5}\right)-1}, \\
\alpha_h &= 170\exp\left(-\frac{V+90}{20}\right), &
\beta_h &= \frac{1000}{1+\exp\left(-\frac{V+42}{10}\right)},\\
\alpha_n &= \frac{-0.1(V+50)}{\exp\left(-\frac{V+50}{10}\right)-1}, &
\beta_n &= 2\exp\left(-\frac{V+90}{80}\right),\\
i_\text{Na} &= (g_\text{Na}+140)(V-E_\text{Na}), &
i_\text{K} &= (g_\text{K1}+g_\text{K2})(V-E_\text{K}),\\
i_\text{leak} &= 75\times\theta_3(V+60), \\
E_\text{Na} &= 40, &
E_\text{K} &= -100,\\
g_\text{Na} &= 400000\times\theta_1m^3h, &
g_\text{K2} &= 1200\times\theta_2n^4,\\
g_\text{K1} &= 1200\times\theta_2\exp\left(-\frac{V+90}{50}\right)+15\exp\left(\frac{V+90}{60}\right), \span \span
\label{equ:noble}
\end{aligned}
\end{equation}

\begin{equation}
\begin{aligned}
\alpha_m &= \frac{-100(V+48)}{\exp\left(-\frac{V+48}{15}\right)-1},&
\beta_m &= \frac{120(V+8)}{\exp\left(\frac{V+8}{5}\right)-1},\\
\alpha_h &= 170\exp\left(-\frac{V+90}{20}\right),&
\beta_h&= \frac{1000}{1+\exp\left(-\frac{V+42}{10}\right)},\\
\alpha_n &= \frac{-0.1(V+50)}{\exp\left(-\frac{V+50}{10}\right)-1},&
\beta_n &= 2\exp\left(-\frac{V+90}{80}\right),\\
g_\text{K1} &= 1200\times\theta_2\exp\left(-\frac{V+90}{50}\right)+15\exp\left(\frac{V+90}{60}\right),\span \span \\
g_\text{K2} &= 1200\times\theta_2n^4,&
g_\text{Na} &= 400000\times\theta_1m^3h,\\
i_\text{Na} &= (g_\text{Na}+140)(V-E_\text{Na}),&
E_\text{Na} &= \frac{RT}{F}\ln\left(\frac{[\text{Na}]_o}{[\text{Na}]_i}\right),\\
i_\text{K} &= (g_\text{K1}+g_\text{K2})(V-E_\text{K}),&
E_\text{K} &= \frac{RT}{F}\ln\left(\frac{[\text{K}]_o}{[\text{K}]_i}\right),\\
i_\text{leak} &= 75\times\theta_3(V+60),&
R &= 8.314,\\
T &= 310,&
F &= 96.485,\\
[\tilde{\text{Na}}]_i &= [\text{Na}]_o\exp\left(-\tilde{E}_\text{Na}\frac{F}{RT}\right),&
[\tilde{\text{K}}]_i &= [\text{K}]_o\exp\left(-\tilde{E}_\text{K}\frac{F}{RT}\right),\\
\tilde{E}_\text{Na} &= 40,&
\tilde{E}_\text{K} &= -100,\\
\diff{V}{t} &= -\frac{i_\text{Na}+i_\text{K}+i_\text{leak}}{12},&
\diff{m}{t} &= \alpha_m(1-m)-\beta_mm,\\
\diff{h}{t} &= \alpha_h(1-h)-\beta_hh,&
\diff{n}{t} &= \alpha_n(1-n)-\beta_nn,\\
\diff{[\text{Na}]_i}{t} &= 2.5\left(-\frac{i_\text{Na}}{1000F} - \frac{[\text{Na}]_i-[\tilde{\text{Na}}]_i}{20}\right),&
[\text{Na}]_o &= 135,\\
\diff{[\text{K}]_i}{t} &= 2.5\left(-\frac{i_\text{K}+i_\text{leak}}{1000F} - \frac{[\text{K}]_i-[\tilde{\text{K}}]_i}{20}\right),&
[\text{K}]_o &= 3.8.
\label{equ:nobleConc}
\end{aligned}
\end{equation}

\section{Appendix: MCMC} \label{sup:mcmc}
\subsection{Data generation}
The data are generated using the parameters $\theta_1,\theta_2,\theta_3=1$. 
The limit cycle is converged over a \qty{10000}{\second} ODE simulation.
From this limit cycle, the period is calculated by the simulated time between two consecutive upcrossings of $V=\qty{-20}{\mV}$. The period is then rounded down to the nearest \qty{1}{\ms}.
A final ODE simulation is then run for this period, aligned to begin at an upcrossing of $V=\qty{-20}{\mV}$ and recorded every \qty{1}{\ms}.
Once the final ODE solution is generated, we add normally distributed iid noise ($\sigma=\qty{2}{\mV}$) to create the synthetic optimization data.

When comparing the converged limit cycles for other parameters, we run the ODE solver from the converged limit cycle, for the period of the synthetic data (rather than the period of this limit cycle). 
Again the ODE solution (membrane voltage) is recorded every \qty{1}{\ms} and aligned to begin at an upcrossing of $V=\qty{-20}{\mV}$.

\subsection{Algorithm}
The adaptive MCMC algorithm is derived from \citet{johnstoneUncertaintyVariabilityModels2016}. The only changes made here are for limit cycle tracking and the full algorithm is shown in Algorithm \ref{alg:mcmc}. The likelihood is calculated as given in Algorithm \ref{alg:ll}.

The prior distribution $\pi(\cdot)$ is the improper uniform distribution $\mathcal{U}(0,\infty)$ for the conductance scale factor parameters and the noise standard deviation prior is given by $\sigma \sim \operatorname{InverseGamma}(\alpha=2, \theta=3)$\footnote{Where the probability density function is parameterized as $f(x;\alpha,\theta)=\theta^\alpha x^{-(\alpha+1)}\exp{(-\theta/x)}/\Gamma(\alpha)$}. 

\SetKwComment{Comment}{/* }{ */}
\begin{algorithm}
\caption{Adaptive MCMC with limit cycle tracking}\label{alg:mcmc}
\KwData{$n_\mathrm{Iterations} > 0$, initial parameters $\theta_0$, standard initial condition $y_0$, log likelihood function $\ell$, observed data $y$}
$a = 1.0$\;
$n_\mathrm{adaption} = \lceil n_\mathrm{Iterations}/10\rceil$ \Comment*[r]{Start adaptive covariance after 10\% of iterations}
$\Sigma = diag(\theta_0/10^6))$\;
$lc_\mathrm{old} = \mathrm{converge}(y_0)$\;
\For{$i\gets1$ \KwTo $n_\mathrm{Iterations}$}{
    $\theta^* \sim \mathcal{N}(\theta,a\Sigma)$\;
    $lc_\mathrm{new}=\mathrm{converge}(lc_\mathrm{old})$\;
    $\alpha = min(1, \exp (\ln(\pi(\theta^*)) + \ell_\mathrm{new} - \ln(\pi(\theta)) - \ell_\mathrm{old}))$\; 
    \eIf{$\alpha>u \sim \mathcal{U}(0,1)$}{
        $\theta = \theta^*$\;
        $lc_\mathrm{old} = lc_\mathrm{new}$\;
        $\ell_\mathrm{old} = \ell_\mathrm{new}$\;
        $\mathrm{accepts}_i=1$\;
    }{
        $\mathrm{accepts}_i=0$
    }
    \If{$i>n_\mathrm{adaption}$}{
        $s = i - n_\mathrm{adaption}$\;
        $\gamma = (s+1)^{-0.6}$\;
        $\Sigma = (1-\gamma)\Sigma + \gamma(x - \theta_0)(x - \theta_0)'$\;
        $\theta_0 = (1-\gamma)\theta_0 + \gamma x$\;
        $a = a\exp(\gamma(accepts_i - 0.25))$\;
    }
}
\end{algorithm}

\begin{algorithm}
\caption{Log likelihood calculation from converged limit cycle}\label{alg:ll}
\KwData{Parameters $p$, a point on the converged limit cycle $c$, noise standard deviation estimate $\sigma$, observed data $y$}
\KwResult{Log likelihood $\ell$}
Solve model until an upcrossing at $V=\qty{-20}{\mV}$, starting from $u_0=c$, with parameters $p$\;
Continue this simulation, recording every \qty{1}{\ms} for the same period as the data, saving as $\hat{y}$\;
$\ell = -\frac{n}{2} \ln(2\pi)-\frac{n}{2}\ln(\sigma^2)-\frac{1}{2\sigma^2} \sum (\hat{y}_i-y_i)^2$\ where $n$ is the number of data points in $y$;
\end{algorithm}

\subsection{Hyperparameters}
The MCMC is run for \qty{40000}~ iterations with the samples from the first 25\% of iterations removed as a burn-in stage. 
The true data generating parameters are $\theta_1,\theta_2,\theta_3=1, \sigma=2$ and the initial parameters are $\theta_1\approx0.993, \theta_2\approx0.996, \theta_3\approx0.973, \sigma=1.5$ with the parameters $\theta_1,\theta_2,\theta_3$ determined as the maximum likelihood estimates through Nelder-Mead optimization in \texttt{Optim.jl} \citep{mogensenOptimMathematicalOptimization2018}. 

The data (including the sampled noise) are identical across all methods.

\subsection{Posteriors and convergence}
\begin{figure}
    \centering
    \includegraphics[width=\textwidth]{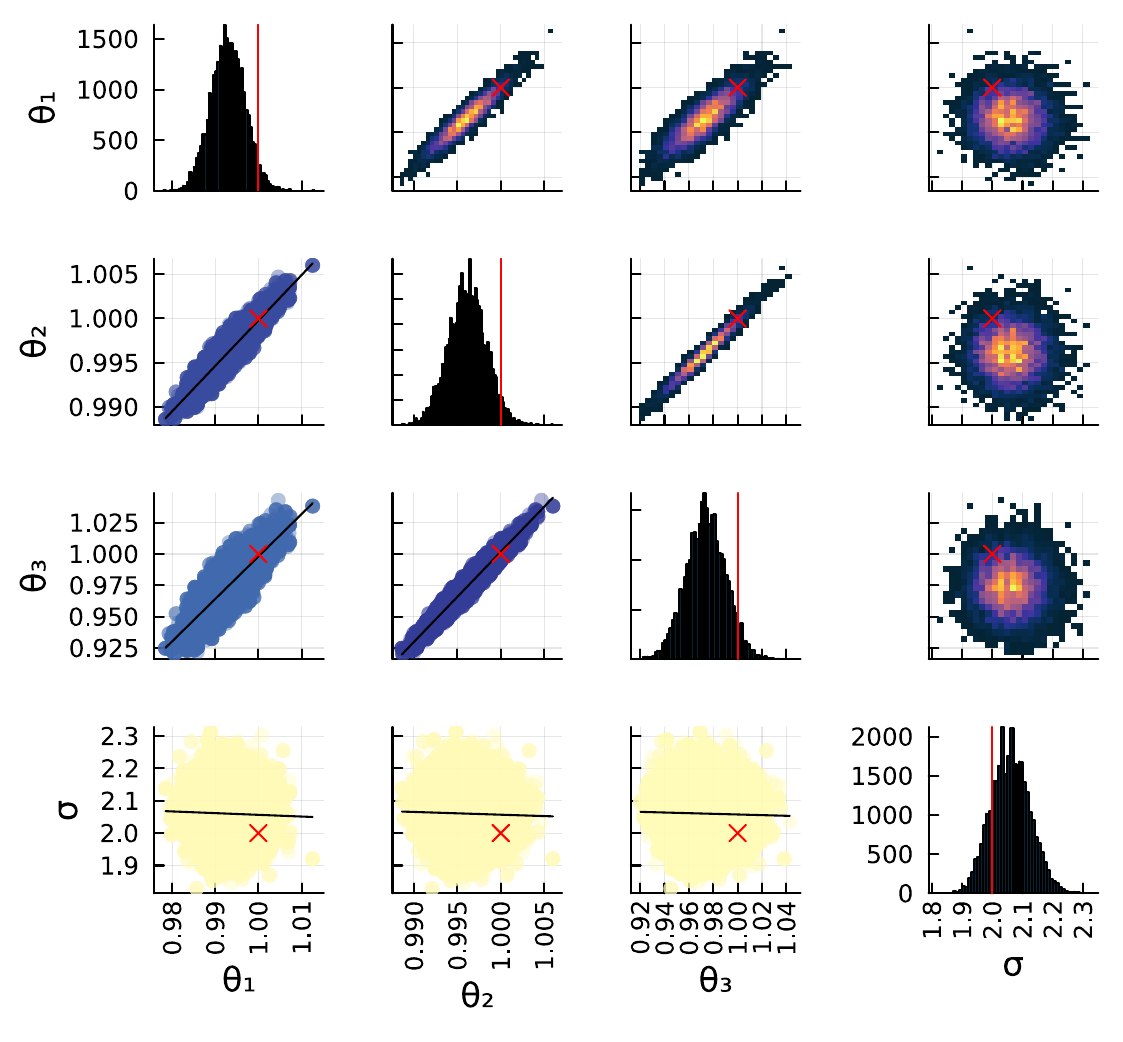}
    \caption{MCMC estimated posterior. The Noble concentration model, Eq.~\eqref{equ:nobleConc}, is fitted to a single noisy converged AP from the data generating parameters $\theta_1=1, \theta_2=1,\theta_3=1$. The method for generating the data and calculating the likelihood are given in Appendix~\ref{sup:mcmc}. The lower triangle shows scatter plots (blue: strong correlation, yellow: weak correlation). The upper triangle shows heatmaps of the bivariate densities. The diagonal contains histograms of the marginal densities. The data generating parameters are given in red. Results shown are from the Standard approach.}
    \label{fig:posteriorStandard}
\end{figure}

\begin{figure}
    \centering
    \includegraphics[width=\textwidth]{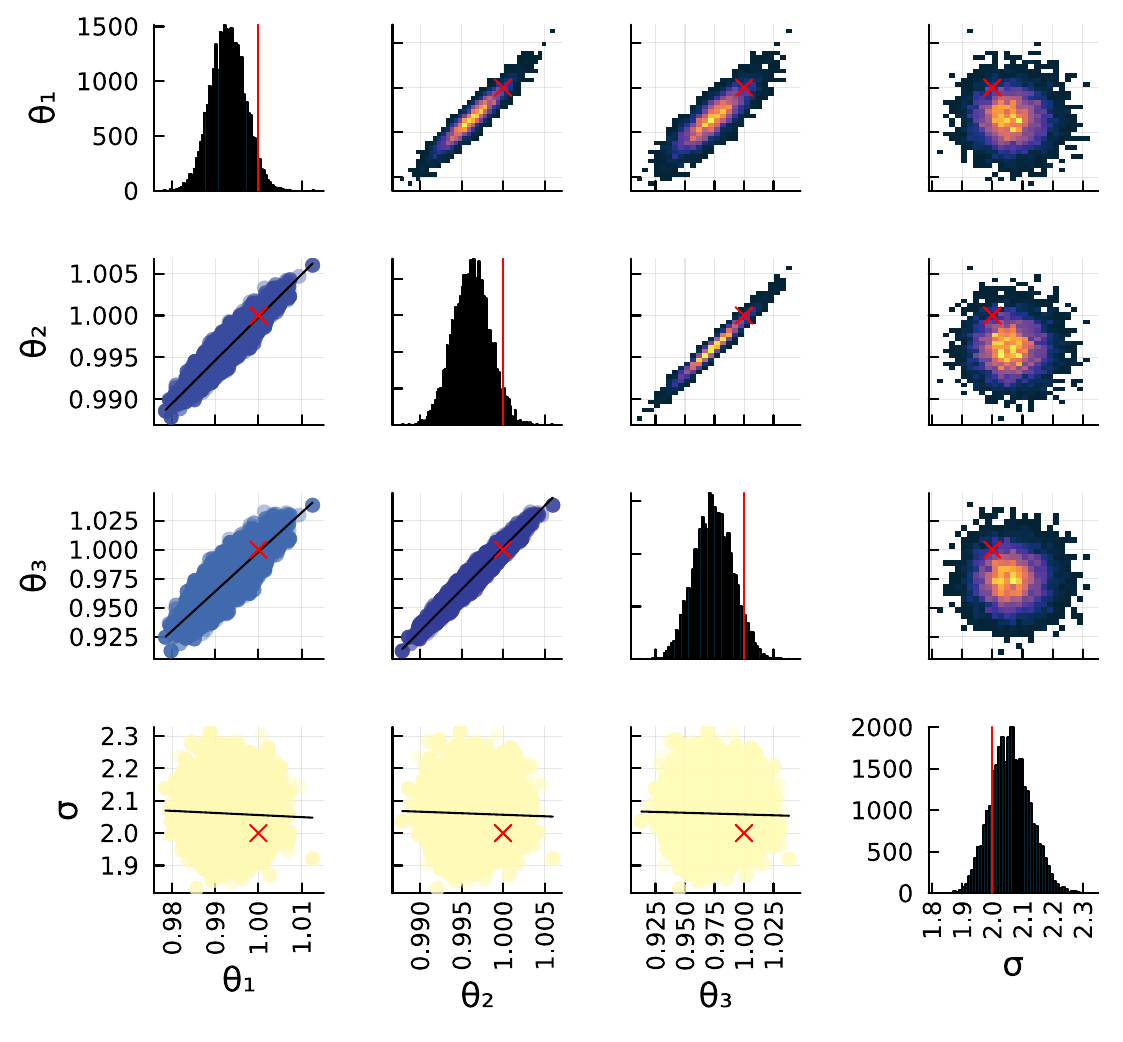}
    \caption{MCMC estimated posterior. The Noble concentration model, Eq.~\eqref{equ:nobleConc}, is fitted to a single noisy converged AP from the data generating parameters $\theta_1=1, \theta_2=1,\theta_3=1$. The method for generating the data and calculating the likelihood are given in Appendix~\ref{sup:mcmc}. The lower triangle shows scatter plots (blue: strong correlation, yellow: weak correlation). The upper triangle shows heatmaps of the bivariate densities. The diagonal contains histograms of the marginal densities. The data generating parameters are given in red. Results shown are from the Tracking approach.}
    \label{fig:posteriorTracking}
\end{figure}

\begin{figure}
    \centering
    \includegraphics[width=\textwidth]{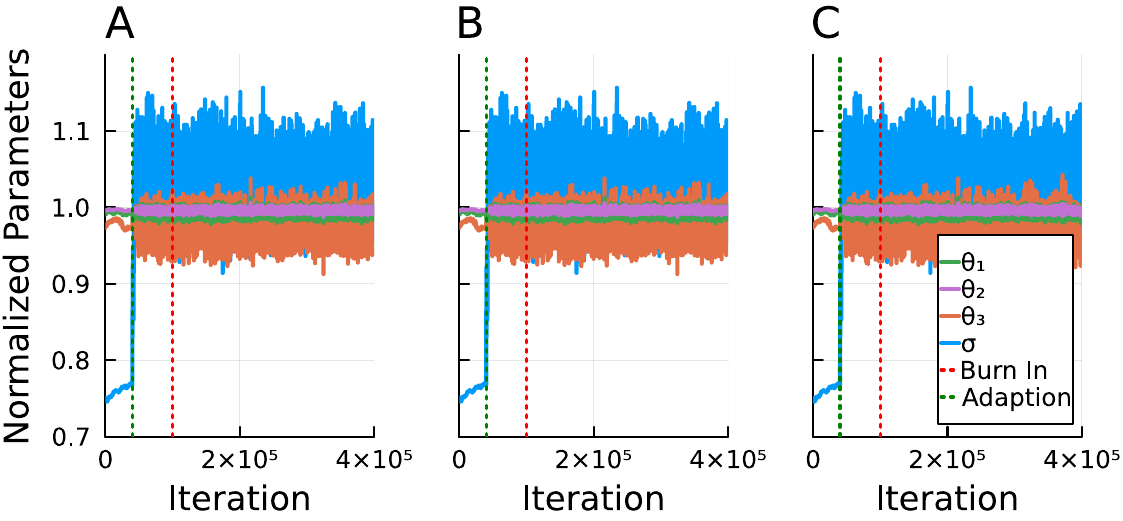}
    \caption{Parameter convergence. Recorded over the MCMC run for each convergence approach. A: Continuation approach, B: Tracking approach, C: Standard approach.}
    \label{fig:paramConv}
\end{figure}

\begin{figure}
    \centering
    \includegraphics[width=\textwidth]{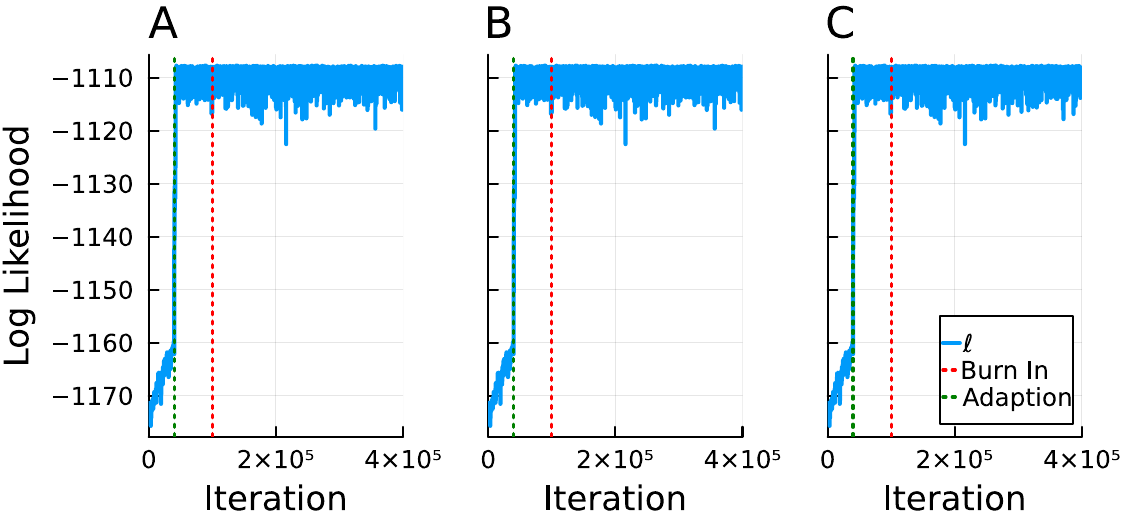}
    \caption{Convergence of the log likelihood. Recorded over the MCMC run for each convergence approach. A: Continuation approach, B: Tracking approach, C: Standard approach.}
    \label{fig:llConv}
\end{figure}

\begin{figure}
    \centering
    \includegraphics[width=\textwidth]{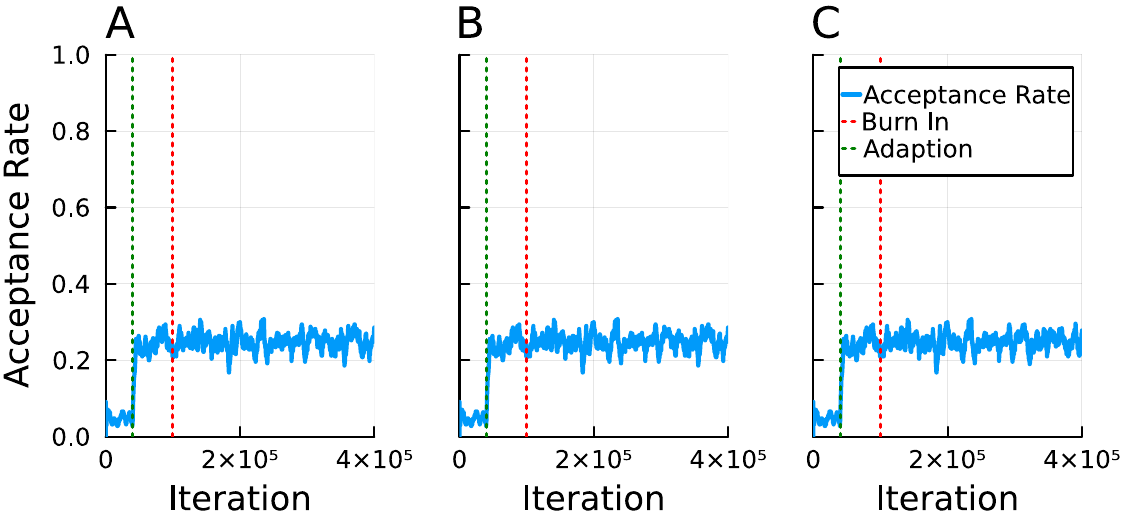}
    \caption{Convergence of the acceptance rate. Averaged over an interval of \qty{100}~ iterations. Recorded over the MCMC run for each convergence approach. A: Continuation approach, B: Tracking approach, C:Standard approach.}
    \label{fig:acceptConv}
\end{figure}

In this section, we provide figures of the posterior distributions for the other two approaches (Figures \ref{fig:posteriorStandard}, \ref{fig:posteriorTracking} for the Standard and Tracking approaches, respectively), parameter convergence (Figure \ref{fig:paramConv}), log likelihood convergence (Figure \ref{fig:llConv}), and acceptance rate convergence (Figure \ref{fig:acceptConv}).

The posteriors are generated after at 25\% burn in phase (which comes after the MCMC has converged).
The posterior distributions across the three different approaches appear identical, demonstrating the approaches behave similarly in terms of accuracy.

The Tracking and Continuation approaches follow the same MCMC chain (follow exactly the same sequence of accepts and rejects), which both differ from the Standard approach after iteration 30275.
At this iteration, the Standard approach identifies a slightly different limit cycle (Standard approach $\ell=-1113.1446$, Continuation approach $\ell=-1113.1460$) and this results in a slightly different acceptance probability (Standard approach $\alpha=0.07555$, Continuation approach $\alpha=0.07541$). 
This small difference results in the Standard approach accepting the proposed parameters and the Continuation approach (and Tracking approach) rejecting the proposed parameters. 
It is likely that using a stricter definition of convergence would result in all approaches identifying the same MCMC chain.

\FloatBarrier

\bibliographystyle{unsrtnat} 
\bibliography{references}

\end{document}